\documentclass[11pt]{article}
\usepackage{endnotes}
\usepackage{revindent}
\begin{document}

\title{Shaping Attitudes Toward Science in an Introductory Astronomy Class}

\author{David Wittman \\University of California, Davis, CA}
\date{\today}
\maketitle

At many universities, astronomy is a popular way for non-science
majors to fulfill a general education requirement. Because
general-education astronomy may be the only college-level science
course taken by these students, it is the last chance to shape the
science attitudes of these future journalists, teachers, politicians,
and voters. Hobson\endnote{A Hobson, ``The Surprising Effectiveness of
  College Scientific Literacy Courses'', Physics Teacher 46, 404
  (2008)} recently reported on research indicating that general
education requirements (often as little as a single class) are
responsible for the higher measured level of science literacy in the
US as compared to Europe.

Science and society will benefit from these students gaining a deeper
understanding of the process of science, rather than focusing
exclusively on its results.  Once outside the university, these
students will find themselves in a society in which two-thirds of
adults (and 38\% of college graduates!) believe that humans were
created ``pretty much in their present form at one time within the
last 10,000 years.''\endnote{The two-thirds figure includes those who
  rated the statement as ``definitely true'' or ``probably true'' in a
  USA Today/Gallup poll published June 7, 2007:
  http://www.usatoday.com/news/politics/2007-06-07-evolution-poll-results\_N.htm?csp=34. Polls
  do not ask specifically about the age of the Earth, but one can take
  this as a rough indicator of public attitudes. The 38\% figure for
  college graduates comes from a meta-analysis of several Gallup polls
  with slightly varying wording: http://www.gallup.com/poll/21811/
  American-Beliefs-Evolution-vs-Bibles-Explanation-Human-Origins.aspx.}
They should be equipped with the tools to refute or at least resist
common anti-science attitudes such as ``we can't know what happened in
the distant past because no one was there to observe it.''  Astronomy,
like evolutionary biology and geology, makes use of fossil evidence,
so instructors have an opportunity to address this issue directly.

If shaping student attitudes is an important goal to us as
instructors, then we should measure those attitudes before and after
instruction, and consider the connection between attitude changes and
classroom activities. I report here on pre- and post-instruction
surveys of science attitudes in my general education astronomy course,
and on classroom activities designed to influence those attitudes. The
first step, well before instruction begins, is to design the attitude
survey.

\section*{Measuring attitudes}

Those who teach introductory physics should be aware that several
attitude assessment instruments are readily available, including
CLASS\endnote{W.K Adams et al., ``A new instrument for measuring
  student beliefs about physics and learning physics: the Colorado
  Learning Attitudes about Science Survey,'' Phys. Rev. ST
  Phys. Educ. Res. 2, 010101 (2006); see also
  http://CLASS.colorado.edu.} (Colorado Learning Attitudes about
Science Survey), MPEX\endnote{Edward F. Redish et al., ``Student
  Expectations in Introductory Physics,'' Am. J. Phys. 66, 212-224
  (1998); see also http://www.physics.umd.edu/perg/expects/index.html}
(Maryland Physics Expectations Survey), and VASS\endnote{I. Halloun \&
  D. Hestenes, ``Views About Science Survey'', Annual Meeting of the
  National Association for Research in Science Teaching. Saint Louis,
  Missouri (1996). ERIC Document No. ED394840. See also
  http://cresmet.asu.edu/prods/vass.shtml.} (Views About Science
Survey). Each of these probes attitudes toward learning, especially
problem-solving, more heavily than attitudes toward science. VASS is
somewhat stronger on the scientific process and the nature of
scientific laws than the other two, and comes with a
taxonomy\endnote{http://cresmet.asu.edu/prods/VASS-P204 Taxonomy.pdf}
which clearly lays out the motivation for each question. For example,
in the epistemology section, the taxonomy states that ``Science is a
coherent body of knowledge about patterns in physical realities (real
systems or phenomena), - rather than a loose collection of particular
empirical facts'' and points to questions 5, 6, and 7 as specifically
probing that issue. In a section labeled ``readiness to learn'' the
taxonomy states that ``science is learnable by anyone willing to make
the effort, - not just by a few talented people'' and points to
question 29 as the relevant probe. Collectively, the statements in the
VASS taxonomy provide an excellent summary of what is missing in the
general public’s attitudes toward science.

One should use an existing survey if at all appropriate, because they
have been extensively field-tested over a range of institutions and
instruction styles. However, none of the above are appropriate for a
general education astronomy course, so I wrote a set of questions
specifically for my course. I started with general science attitudes
such as those mentioned above in the discussion of VASS.  For example,
students are asked to rate the statement ``Science is largely about
memorizing facts'' on a Likert scale (strongly disagree to strongly
agree on a scale of 1-5 or $a-e$ on a scantron form). I included
several questions related to this attitude, because in a large lecture
hall, in a nonmathematical course with no lab component (none of these
parameters is under my control), it would be easy to miss the process
of science and slip into merely describing the results.  Another
example, motivated in part by science coverage in the media which
celebrates ``discoveries'' without considering the surrounding
framework: ``Scientists do experiments: (a) on what interests them the
most; (e) on things that could verify or falsify theories.'' This is a
forced-choice type of question, in which students must choose between
the two options on a five-letter scale, with $a$ indicating complete
agreement with the first option and $e$ indicating complete agreement
with the second option.

Because this is an astronomy course, I added items relating to
reconstructing the past and our place in the universe, for example
``We can't know what happened in the distant past because we don’t
have any observations of that time'' and ``We live in a changing
universe.''  A few questions are somewhat factual in nature, but
relate strongly to attitudes about our place in the universe,
e.g. ``The Sun is a special star.''

The list of nineteen questions on science attitudes appears below
along with the results; ; a copy of the survey also appears at the end
of this preprint and is available for anyone to use freely.  I gave
the survey at the start of the first day of class and at the end of
the final day. Surveys are more conveniently delivered online, but
instructors should weigh this convenience against the likelihood of
getting online responses from a biased sample of more conscientious
students (or offer extra credit and issue several email
reminders). Because there is a good deal of dropping and adding
courses in the first two weeks of the term, I asked students to write
their student identification numbers on the survey so that I could
rigorously match pre- and post-instruction. I assured them that their
attitudes would not be studied individually, but writing their ID may
nevertheless have put them on better behavior compared to a completely
anonymous survey.

\section*{Shaping attitudes}

If writing the attitude survey helps clarify the goals, clear goals in
turn guide the construction of classroom activities to further those
goals. Activities and case studies are more likely to influence
attitudes than simply lecturing about the nature of science, and
making the nature of science a recurring theme is likely to have more
impact than a single first-day activity. While I can't prove a causal
relationship between these activities and attitude changes, I offer a
few examples which instructors may find useful.

The first week, I give the students Ob-scertainers-—-little black
boxes whose internal structures they must deduce from the sound of a
ball bearing rolling around inside. They work in groups of two to form
hypotheses, then groups double up to compare hypotheses and conduct
additional tests. At the end, when I am no doubt expected to reveal
the correct answer, I tell them that nature never reveals the correct
answer to scientists.\endnote{Credit for this idea goes to Steve Shawl
  of Kansas University.} All we can do is obtain more data and refine
our hypotheses, so I ask the students to suggest additional methods of
doing so. They offered many creative and enthusiastic responses to
this open-ended question.  

The energy source of the Sun and the age of the solar system make a
nice case study. In the late 1800’s fusion was not yet known, and the
largest energy source astronomers knew of was Kelvin-Helmholtz
contraction, which could power the Sun for about 25 million years. Yet
geologists were confident from the study of rocks that the Earth was
of order 1 billion years old. What would you decide if this was all
the information you had? Is the statement that ``astronomers don't
know of an energy source that can power the Sun for 1 billion years''
enough to discredit the geological evidence?  On the other hand,
should we just assume from the geological evidence that there must be
a source of power orders of magnitude beyond what is currently known?
This relates not only to the tentative nature of scientific
conclusions, but also the interconnectedness of all branches of
science, and the ability to recognize and mentally process the
existence of large gaps in our knowledge. The fact that astronomers
later radically changed their views of the Sun’s power source is an
indication of science's strength, not its inefficacy, a point that
should serve students well when reading modern-day press releases
which tend to overly emphasize challenges to established scientific
views.

Finding hands-on activities that work well in a large lecture hall can
be a challenge.  Particularly well-suited is a demonstration of why
photons take of order a million years to travel from the core of the
Sun to its surface. A balloon is placed in the center of the audience
and students are first told to pass it forward; it quickly leaves the
audience. The balloon is reset in the center and this time students
are told to pass it in random directions. It takes much longer to
reach the edge of the audience. (This is one of those rare activities
where the larger the lecture hall, the better.) I use the activity not
only to demonstrate the idea, but to let the students participate in
making and testing models and lead them to understand that scientists
do the same thing on a more mathematical level.

Another good case study is the Shapley-Curtis debate over the scale of
the universe: Are the spiral nebulae small objects within our own
Galaxy, or are they large, distant ``island universes'' unto
themselves?  At the time of the formal debate in 1920, the evidence
was conflicting, but—--and this is key to the nature of science—--when
solid new evidence was presented, the debate was over. Furthermore, an
important piece of data against the ``island universe'' interpretation
was later shown to have been incorrectly measured. This provides an
excellent opportunity to discuss how scientists reach conclusions. I
poll the class after presenting the original evidence, then after
retracting some of the evidence, etc. This illustrates the conditional
nature of scientific conclusions.

\section*{Results}

Post-instruction, I first matched the pre and post tests to yield a
set of $n=63$ students who had taken both. Next, I inverted the scores
on questions for which the “correct” attitude was 1 or $a$ rather than
5 or $e$, so that a higher number is always considered better. In
measuring the learning of {\it content}, it is standard to define the
{\it gain} as posttest score minus pretest score, and {\it normalized
  gain} as (posttest-pretest)/(maximum-pretest), in other words, how
much the students learned as a fraction of what they didn't already
know.\endnote{R. R. Hake, ``Interactive-engagement vs traditional
  methods: A six-thousand-student survey of mechanics test data for
  introductory physics courses,'' Am.J.Phys. 66, 64 (1998)} I adopt
these definitions here, with the caveat that this survey deals with
{\it attitude changes} which is not the same as learning content. For
example, expert attitudes will not always be 5.0 because of some
difficult forced choices, and should be determined by giving the same
survey to professional scientists. This is a clear advantage of using
an existing instrument such as VASS, which has been extensively
field-tested in different populations, {\it if} it is appropriate to
the course. However, the lack of an appropriate field-tested survey
should not prevent instructors who care about attitudes from trying to
measure them. The results cannot be interpreted as rigorously, but the
process can lead to better teaching and learning.

I list the results below in descending order of normalized gain,
expressed here as a percentage, in a recent term. In parentheses, I
list the pre- and post-instruction average attitude.

\begin{invertedparagraphs}

57\% (4.44, 4.76) A bit of information is considered scientific when:
(a) it is supported by evidence; (e) it is supported by eminent
scientists.

52\% (3.30, 4.19) Most of the atoms in your body were once inside a star.

50\% (4.14, 4.57) If a scientist publishes an interesting new theory,
other scientists usually: (a) compete to prove it wrong; (e) ignore it
because it is not their theory.

49\% (3.47, 4.22) We can't know what happened in the distant past
because we don't have any observations of that time.

41\% (4.24, 4.56) We live in a changing universe.

39\% (4.64, 4.78) If a scientist publishes an interesting experimental
result which is in error, the error is usually detected by: (a) no
one; (e) other scientists attempting to reproduce the result.

31\% (3.56, 4.02) Science is largely about memorizing equations.

29\% (2.87, 3.51) The Sun is a special star.

27\% (4.21, 4.43) Science affects everyone's life.

26\% (3.29, 3.74) We can't know what happened in the distant past
because we can't do controlled experiments on it.

24\% (4.54, 4.65) An experimental result: (a) should be trusted
because it is scientific; (e) should be questioned as to how it fits
in with other experimental results.

21\% (3.43, 3.76) Science is largely about memorizing facts.

20\% (4.06, 4.25) Scientists who created successful theories did it
by: (a) having exceptional intelligence; (e) looking for patterns in
data.

14\% (3.57, 3.78) Science is largely about discovering patterns.

10\% (3.60, 3.75) Anyone can understand the major themes of science if they try.

9\% (2.87, 3.08) Astronomy is difficult because the physical laws that
apply on Earth do not necessarily apply everywhere.

0\% (3.68, 3.68) I want my tax money to help support science.

-3\% (3.70, 3.65) Science is fun.

-9\% (3.43, 3.29) Scientists do experiments: (a) on what interests them
the most; (e) on things that could verify or falsify theories.
\end{invertedparagraphs}

The typical per-question statistical uncertainty in gain
(posttest-pretest) is 0.12 points. Thus the two declines are not
statistically significant. Several attitudes improved by many times
the statistical uncertainty, with a normalized gain of $\sim$50\%,
comparable to the typical normalized gain for {\it subject matter} in
effective courses.$^9$ Many other attitudes individually showed a gain
of only a few times the statistical uncertainty, but collectively they
represent a significant shift.

\section*{Summary and conclusions}

It {\it is} possible to influence student attitudes for the
better. Several previous studies have shown that attitudes tend to
decline or remain unchanged after introductory physics and astronomy
courses,$^{4,}$\endnote{M. Zeilik et al, ``Conceptual
  astronomy. II. Replicating conceptual gains, probing attitude
  changes across three semesters'', Am.J.Phys. 67, 923 (1999)} with
just a few small-scale counterexamples.\endnote{A. Elby, “Helping
  physics students learn how to learn,” Am. J. Phys. 69, S54 (2001)}$^,$\endnote{
  J. Marx and K. Cummings, ``What Factors Really Influence Shifts in
  Students’ Attitudes and Expectations in an Introductory Physics
  Course?'', AIP Conference Proceedings 883, 101 (2007)} However,
these studies have focused more on attitudes toward
learning. Attitudes toward science may be less deeply ingrained.

The results here are promising, but with no control group the efficacy
of the instructional approach cannot be rigorously established.  In
retrospect, the process of preparing the attitude survey was key to
the entire venture, as it forced me to reflect on and organize my
objectives with respect to student attitudes.  I encourage other
instructors to engage in this process and see where it leads.

\theendnotes

\newpage

\begin{verbatim}
These questions are designed to measure the incoming attitudes of the
class.  Please answer honestly.  You will not be identified
individually.  (We ask for your student ID on the scantron only so
that we can look at attitudes of different groups.)  Please do not
mark the questions, so that the paper can be reused.

For questions which are simple statements, such as number 1, respond as follows:
1: strongly disagree
2: somewhat disagree
3: neither agree nor disagree
4: somewhat agree
5: strongly agree

For questions which have two alternatives, such as number 2, respond as follows:
1: strongly agree with alternative a rather than e
2: somewhat agree with alternative a rather than e
3: agree about the same with a or e
4: somewhat agree with alternative e rather than a
5: strongly agree with alternative e rather than a

------------------------------------------------------------------------------

1. Science is largely about memorizing facts.

2. Scientists do experiments:
(a) on what interests them the most
(e) on things that could verify or falsify theories

3. Scientists who created successful theories did it by
(a) having exceptional intelligence
(e) looking for patterns in data

4. Science is largely about memorizing equations. 

5. A bit of information is considered scientific when: 
(a) it is supported by evidence
(e) it is supported by eminent scientists

6. Anyone can understand the major themes of science if they try.

7. Science is fun.

8. An experimental result:
(a) should be trusted because it is scientific.
(e) should be questioned as to how it fits in with other experimental results.

9. Science affects everyone's life.

10. Science is largely about discovering patterns.

11. If a scientist publishes an interesting experimental result which is in 
error, the error is usually detected by: 
(a) no one 
(e) other scientists attempting to reproduce the result

12. If a scientist publishes an interesting new theory, other scientists usually:
(a) compete to prove it wrong
(e) ignore it because it is not their theory

13. We live in a changing universe.

14. The Sun is a special star.

15. Astronomy is difficult because the physical laws that apply on
Earth do not necessarily apply everywhere.

16. Most of the atoms in your body were once inside a star.

17. I want my tax money to help support science.

18. We can't know what happened in the distant past because we don't
have any observations of that time.

19. We can't know what happened in the distant past because we can't do
controlled experiments on it.

Thank you for completing this survey!

\end{verbatim}

\end{document}